\RequirePackage{fix-cm}
\documentclass[twocolumn,epjc3]{svjour3}  
\smartqed  % flush right qed marks, e.g. at end of proof
\RequirePackage{graphicx}
\journalname{Eur. Phys. J. C}
\begin{document}

\title{Systematic study of the Chiral Magnetic Effect with the AVFD model at LHC energies%\thanksref{t1}
}
%\subtitle{Do you have a subtitle?\\ If so, write it here}

%\titlerunning{Short form of title}        % if too long for running head

\author{Panos Christakoglou\thanksref{e1,addr1}
        \and
        Shi Qiu\thanksref{e2,addr1} %etc.
        \and
        Joey Staa\thanksref{e3,addr1} %etc.
}

%\thankstext{t1}{Grants or other notes
%about the article that should go on the front page should be
%placed here. General acknowledgments should be placed at the end of the article.
\thankstext{e1}{e-mail: Panos.Christakoglou@nikhef.nl}
\thankstext{e2}{e-mail: Shi.Qiu@nikhef.nl}
\thankstext{e3}{e-mail: Joey.Staa@nikhef.nl}

%\authorrunning{Short form of author list} % if too long for running head

\institute{Nikhef \label{addr1}
           %\and
           %Utrecht University \label{addr2}
}

\date{Received: date / Accepted: date}
% The correct dates will be entered by the editor

\maketitle

\begin{abstract}
We present a systematic study of the correlators used 
experimentally to probe the Chiral Magnetic Effect (CME) 
using the Anomalous Viscous Fluid Dynamics (AVFD) model in Pb--Pb 
and Xe--Xe collisions at LHC energies. We find a parametrization 
that describes the dependence of these correlators on 
the value of the axial current density ($n_5/\mathrm{s}$), 
which dictates the CME signal, and on the parameter that 
governs the background in these measurements i.e., the 
percentage of local charge conservation (LCC) within an 
event. This allows to deduce the values of $n_5/\mathrm{s}$ 
and the LCC percentage that provide a quantitative description 
of the centrality dependence of the experimental measurements. 
We find that the results in Xe--Xe collisions at 
$\sqrt{s_{\mathrm{NN}}} = 5.44$~TeV are consistent with 
a background only scenario. On the other hand, the model 
needs a significant non-zero value of $n_5/\mathrm{s}$ to 
match the measurements in Pb--Pb collisions at 
$\sqrt{s_{\mathrm{NN}}} = 5.02$~TeV.
% \PACS{PACS code1 \and PACS code2 \and more}
% \subclass{MSC code1 \and MSC code2 \and more}
\end{abstract}

%________________________________________________%
%________________________________________________%
%________________________________________________%
\section{Introduction}
\label{Sec:Introduction}
Collisions between heavy ions accelerated at ultra-relativistic 
energies provide the necessary conditions to form a deconfined 
state of matter, the Quark Gluon Plasma~\cite{Karsch:2003jg}. 
In this phase, the fundamental constituents of quantum 
chromodynamics (QCD), the quarks and gluons, are not anymore 
confined inside their usual hadronic bags. The transition to 
a QGP from normal hadronic matter is expected to take place 
at a temperature of about 155~MeV, and an energy density of 
about 0.5 GeV/fm$^3$, according to lattice QCD calculations 
\cite{Bazavov:2009zn,Bazavov:2011nk,Borsanyi:2010cj}. These 
conditions can be reached in collisions between Pb ions at the 
Large Hadron Collider 
(LHC)~\cite{Chatrchyan:2012mb,Adam:2015lda,Adam:2016thv}.

Heavy ion collisions also provide the possibility to study novel 
QCD phenomena that are otherwise not accessible experimentally. 
One characteristic example is related to local parity (P) as well as charge conjugation and parity (CP) symmetry violation in strong 
interactions. The possibility to observe parity violation in the 
strong interaction using relativistic heavy-ion collisions has 
been discussed in~\cite{Lee:1973iz,Lee:1974ma,Morley:1983wr} and 
was further 
reviewed in~\cite{Kharzeev:1998kz,Kharzeev:1999cz,Kharzeev:2015kna,Kharzeev:2007tn,Kharzeev:2007jp,Kharzeev:2015znc,Li:2020dwr,Kharzeev:2020jxw}. 
In QCD, this symmetry violation originates from the interaction 
between the chiral fermions of the theory and topologically 
non-trivial gluonic fields that induce net-chirality. In the 
presence of a strong magnetic field, such as the one created in 
peripheral heavy ion collisions with a magnitude of around 
$10^{15}$~Tesla~\cite{Skokov:2009qp,Bzdak:2011yy,Deng:2012pc}, 
these interactions lead to an asymmetry between left and right-handed 
quarks. The generated net-chirality, in turns, leads to an excess 
of positively and negatively charged particles moving in opposite 
directions relative to the system's symmetry plane. This introduces 
an electromagnetic current and the creation of an electric dipole 
moment of QCD matter. The experimental search for these effects 
has intensified recently, following the realisation that the 
subsequent creation of charged hadrons results in an experimentally 
accessible magnitude of charge separation along the direction of 
this magnetic field, and perpendicular to the symmetry plane. 
This phenomenon is called the Chiral Magnetic Effect (CME)~\cite{Fukushima:2008xe} 
and its existence was recently reported in semi-metals like zirconium 
pentatelluride ($ZrTe_5$)~\cite{Li:2014bha}.

Early enough it was realised that a way to probe these effects 
is to rely on measuring two-particle azimuthal correlations 
relative to the reaction plane ($\Psi_{\mathrm{RP}}$)~\cite{Voloshin:2004vk}, 
the plane defined by the impact parameter and the beam axis. 
Since then, intensive experimental efforts have been made to 
identify unambiguously signals of the CME. The first measurements 
using this approach were reported by the STAR Collaboration 
in Au--Au collisions at $\sqrt{s_{\mathrm{NN}}} = 0.2$~TeV~\cite{Abelev:2009ac,Abelev:2009ad} 
and were consistent with initial expectations for a charge 
separation relative to the reaction plane due to the CME. Soon 
after, the first results from the LHC in Pb--Pb collisions at 
$\sqrt{s_{\mathrm{NN}}} = 2.76$~TeV~were reported and showed a 
quantitatively similar effect~\cite{Abelev:2012pa}. This agreement 
between the results is up until this moment hard to comprehend 
considering the  differences in the centre-of-mass energy and 
consequently in the multiplicity density~\cite{Aamodt:2010cz}. 
In addition, the magnetic field and the way it evolves is, in 
principle, different between the two 
energies~\cite{Skokov:2009qp,Bzdak:2011yy,Deng:2012pc}. Overall, 
this agreement hinted at the dominant role of background effects 
in both measurements. These background effects were, in parallel, 
identified as coming from local charge conservation coupled 
to the anisotropic expansion of the system in non-central 
collisions~\cite{Schlichting:2010qia,Pratt:2010zn}. The field 
turned its focus to finding a way to constrain and quantify the 
background and the CME contribution to such measurements. 

In Ref.~\cite{Acharya:2017fau}, the ALICE Collaboration presented 
the first ever upper limit of 26--33$\%$ at $95\%$ confidence level 
for the CME contribution, using an Event Shape Engineering (ESE) 
technique~\cite{Schukraft:2012ah}. In parallel, new measurements 
of the STAR Collaboration in Au--Au collisions at a centre-of-mass 
energy $\sqrt{s_{\mathrm{NN}}} = 200$~GeV~\cite{Adamczyk:2013kcb,Adamczyk:2013hsi,Adamczyk:2014mzf} 
as well as results obtained from the analysis of data collected 
from the beam energy scan at $\sqrt{s_{\mathrm{NN}}} =$7.7, 11.5, 
19.6, 27, 39 and 62.4~GeV~\cite{Adamczyk:2014mzf} were still qualitatively 
consistent with expectations from parity violating effects 
in heavy ion collisions. To study background effects the 
CMS~\cite{Khachatryan:2016got} and the STAR~\cite{STAR:2019xzd} 
collaborations studied charge dependent correlations in both p--Pb 
collisions at $\sqrt{s_{\mathrm{NN}}} = 5.02$~TeV and in p--Au and 
d--Au collisions at $\sqrt{s_{\mathrm{NN}}} = 0.2$~TeV, respectively. 
Both results illustrate that these correlations are similar to 
those measured in heavy-ion collisions. The authors concluded that these findings could have important 
implications for the interpretation of the heavy-ion data since it 
is expected that the results in these ``small'' systems are 
dominated by background effects. However these latter studies are lacking 
a quantitative estimate of the reaction plane independent 
background~\cite{Abelev:2009ac,Abelev:2009ad} and therefore should 
not be used to extract a definite conclusion. Finally, the ALICE 
Collaboration recently reported their updated upper limits of 15--18$\%$ 
at 95$\%$ confidence level for the centrality interval 0--40$\%$ by 
studying charge dependent correlations relative to the third order 
symmetry plane ~\cite{Acharya:2020rlz}. Overall, the extraction of 
the CME signal has been exceptionally challenging.

In this article we follow a different approach by performing a 
systematic study of the correlators used in CME searches for 
Pb--Pb and Xe--Xe collisions at $\sqrt{s_{\mathrm{NN}}} = 5.02$~TeV 
(for Pb ions)~\cite{Abelev:2012pa,Acharya:2020rlz} and at 
$\sqrt{s_{\mathrm{NN}}} = 5.44$~TeV (for Xe ions)~\cite{Aziz:2020nia} 
with the Anomalous-Viscous Fluid Dynamics (AVFD) framework~\cite{Shi:2017cpu,Jiang:2016wve,Shi:2019wzi}. 
This is a state-of-the-art model that describes the initial state 
of the collision using a Glauber prescription, and accounts for the 
development of the early stage electromagnetic fields as well as for 
the propagation of anomalous fermion currents. The expanding medium 
is treated via a 2+1 dimensional viscous hydrodynamics (VISH2+1) 
code which is coupled to a hadron cascade model 
(UrQMD)~\cite{Bass:1998ca}. The goal of this study is to extract 
the relevant values that govern the CME signal and the background 
in the AVFD model that will allow for a quantitative description 
of the centrality dependence of the charged dependent correlations 
measured in various colliding systems and energies at the LHC. 

The article is organised as follows: Section~\ref{Sec:Observables} 
presents the main observables, followed by a discussion on how the 
model is calibrated in Section~\ref{Sec:Model}. The main results are 
presented in Section~\ref{Sec:Results}. The article concludes with 
a summary.

\section{Experimental observables}
\label{Sec:Observables}
%The gluonic fields responsible for the parity violating effects carry 
%net-chirality with different sign in every event and thus averaged 
%over many events they would not yield a finite expectation value for 
%any P--odd variable. This makes the observation of the CME possible 
%only via P--even observables, expressed in terms of two- and multi-particle 
%correlations that are frequently used in studies of the azimuthal 
%anisotropy~\cite{Voloshin:1994mz}. In these latter studies, the azimuthal 
%distribution of produced particles is expressed by a Fourier transform 
%according to:

%\begin{equation}
%\frac{dN}{d\varphi} \approx 1 + 2\sum_{n} v_n \cos[n(\varphi - \Psi_n)]
%\label{Eq:Fourier}
%\end{equation}

%\noindent where $N$ is the number of particles, $\varphi$ is the 
%azimuthal angle of the particle, $\Psi_n$ is the n-th order symmetry 
%plane of the system created after the collision, and $v_n$ are the 
%corresponding flow coefficients ($v_1$: directed flow, $v_2$: elliptic 
%flow, $v_3$: triangular flow etc.). 

A way to probe the parity violating effects is by introducing P-odd  
coefficients $a_{n,\alpha}$ in the Fourier series frequently used in 
studies of azimuthal anisotropy~\cite{Voloshin:1994mz}. This leads to 
the expression 

\begin{equation}
\frac{dN}{d\varphi} \approx 1 + 2\sum_{n} \Big[v_n \cos[n(\varphi - \Psi_n)] + a_n \sin[n(\varphi - \Psi_n)]\Big]
\label{Eq:Fourier}
\end{equation}

\noindent where $N$ is the number of particles, $\varphi$ is the 
azimuthal angle of the particle and $v_n$ are the corresponding 
flow coefficients ($v_1$: directed flow, $v_2$: elliptic flow, 
$v_3$: triangular flow etc.). The n-th order symmetry plane of 
the system, $\Psi_n$, is introduced to take into account that 
the overlap region of the colliding nuclei exhibits an irregular
shape~\cite{Manly:2005zy,Bhalerao:2006tp,Alver:2008zza,Alver:2010gr,Alver:2010dn}. This originates from the initial density profile of 
nucleons participating in the collision, which is not isotropic 
and differs from one event to the other. In case of a smooth 
distribution of matter produced in the overlap zone, the angle 
$\Psi_{n}$ coincides with that of the reaction plane, 
$\Psi_{\rm RP}$. In Eq.~\ref{Eq:Fourier}, $a_1$ is the leading 
order P-odd term that reflects the magnitude, while higher 
harmonics (i.e. $a_2$ and above) represent the specific shape 
of the CME signal.

In Ref.~\cite{Voloshin:2004vk}, Voloshin proposed that the 
leading order P-odd coefficient can be probed through the 
study of charge-dependent two-particle correlations relative 
to the reaction plane $\Psi_{\rm RP}$. In particular, the 
expression discussed is of the form 
$\langle \cos(\varphi_{\alpha} + \varphi_{\beta} - 2\Psi_{\rm RP}) \rangle$, where $\alpha$ and $\beta$ denote particles with 
the same or opposite charge. This expression can probe 
correlations between the leading P-odd terms for different 
charge combinations $\langle a_{1,\alpha} a_{1,\beta} \rangle$. 
This can be seen if one expands the correlator using 
Eq.~\ref{Eq:Fourier} according to

\[\langle \cos(\varphi_{\alpha} + \varphi_{\beta} - 2\Psi_{\rm RP}) \rangle = \] 

\[ \langle \cos\big[(\varphi_{\alpha} - \Psi_{\rm RP}) +
(\varphi_{\beta} - \Psi_{\rm RP})\big] \rangle = 
\langle \cos(\Delta \varphi_{\alpha} + \Delta \varphi_{\beta}) \rangle = \]

\[\langle \cos \Delta \varphi_{\alpha} \cos \Delta \varphi_{\beta} \rangle - 
\langle \sin \Delta \varphi_{\alpha} \sin \Delta \varphi_{\beta} \rangle = 
\]

\begin{equation}
\langle v_{1,\alpha}v_{1,\beta} \rangle + \mathrm{B_{in}} - 
\langle a_{1,\alpha} a_{1,\beta}\rangle - \mathrm{B_{out}},  
% \langle v_{1,\alpha}v_{1,\beta} \rangle + \mathrm{B_{in}} - 
% \langle a_{1,\alpha} a_{1,\beta} \rangle -
% \mathrm{B_{out}}
\label{Eq:3ParticleCorrelator}
\end{equation}
%In Eq.~\ref{Eq:3ParticleCorrelator}, 
where $\mathrm{B_{in}}$ and $\mathrm{B_{out}}$ represent the 
parity-conserving correlations projected onto the in- and 
out-of-plane directions~\cite{Voloshin:2004vk}. The terms 
$\langle \cos \Delta \varphi_{\alpha} \cos \Delta \varphi_{\beta}\rangle$
and $\langle \sin \Delta \varphi_{\alpha}\sin \Delta\varphi_{\beta}\rangle$
in Eq.~\ref{Eq:3ParticleCorrelator} quantify the correlations 
with respect to the in- and out-of-plane directions, respectively. 
The term $\langle v_{1,\alpha}v_{1,\beta} \rangle$, i.e. the product 
of the first order Fourier harmonics or directed flow, is expected 
to have negligible charge dependence in the mid-rapidity 
region~\cite{Gursoy:2018yai}. In addition, for a symmetric collision 
system the average directed flow at mid-rapidity is zero.
% One of the advantages of using Eq.~\ref{Eq:3ParticleCorrelator} is
% that since it is constructed as the difference between
% $\mathrm{B_{in}}$ and $\mathrm{B_{out}}$ it significantly suppresses 
% non-flow correlations.

A generalised form of Eq.~\ref{Eq:3ParticleCorrelator},  
describing also higher harmonics, is given by the mixed-harmonics 
correlations and reads 
\begin{eqnarray}
\label{eq:moments}
\gamma_{\mathrm{m,n}} = \langle \cos(\mathrm{m}\varphi_{\alpha} +
\mathrm{n}\varphi_{\beta} - \mathrm{(m + n)}\Psi_{\mathrm{|m+n|}})\rangle , 
\label{Eq:Generalised3ParticleCorrelator}
\end{eqnarray}

\noindent where $\mathrm{m}$ and $\mathrm{n}$ are integers. 
Setting $\mathrm{m} = 1$ and $\mathrm{n} = 1$ (i.e. $\gamma_{1,1}$) 
leads to Eq.~\ref{Eq:3ParticleCorrelator}. 

In order to independently evaluate the contributions from 
correlations in- and out-of-plane, one can also measure a 
two-particle correlator of the form

\[ \langle \cos(\varphi_{\alpha} - \varphi_{\beta}) \rangle = \]

\[ \langle \cos\big[(\varphi_{\alpha} - \Psi_{\rm RP}) -
(\varphi_{\beta} - \Psi_{\rm RP})\big] \rangle =  
\langle \cos(\Delta \varphi_{\alpha} - \Delta \varphi_{\beta}) \rangle= \] 

\[\langle \cos \Delta \varphi_{\alpha} \cos \Delta \varphi_{\beta} \rangle +
\langle \sin \Delta \varphi_{\alpha} \sin \Delta \varphi_{\beta} \rangle = 
\]

\begin{equation}
\langle v_{1,\alpha}v_{1,\beta} \rangle + \mathrm{B_{in}} + 
\langle a_{1,\alpha} a_{1,\beta}\rangle + \mathrm{B_{out}},
\label{Eq:2ParticleCorrelator}
\end{equation}

\noindent This provides access to the two-particle correlations 
without any dependence on the symmetry plane angle which can be 
generalised according to

\begin{eqnarray}
\delta_\mathrm{m} = \langle \cos[\mathrm{m}(\varphi_{\alpha} -
\varphi_{\beta})] \rangle .
\label{Eq:Generalised2ParticleCorrelator}
\end{eqnarray}
\noindent This correlator, owing to its construction, is affected 
if not dominated by background contributions. Charge-dependent 
results for $\delta_1$, together with the relevant measurements of 
$\gamma_{1,1}$ were first reported in Ref.~\cite{Abelev:2012pa} and 
made it possible to separately quantify the magnitude of correlations 
in- and out-of-plane.

\section{Model calibration and parametrisation}
\label{Sec:Model}
The goal of this study is to extract the values that control 
the CME signal and the background in the AVFD model that will 
allow for a quantitative simultaneous description of the centrality 
dependence of the charged dependent correlations, i.e. 
$\Delta \delta_1$ and $\Delta \gamma_{1,1}$ measured in Pb--Pb 
collisions at 
$\sqrt{s_{\mathrm{NN}}} = 5.02$~TeV~\cite{Abelev:2012pa,Acharya:2020rlz} 
and in Xe--Xe collisions at $\sqrt{s_{\mathrm{NN}}} = 5.44$~TeV~\cite{Aziz:2020nia}. 
Here, $\Delta \delta_1$ and $\Delta \gamma_{1,1}$ denote the difference 
of $\delta_1$ and $\gamma_{1,1}$ between opposite- and same-sign pairs. 
Within the AVFD framework, the CME signal is controlled by the axial 
current density $n_5/\mathrm{s}$ which dictates the imbalance between 
right- and left-handed fermions induced in the initial stage of each 
event. The parameter that governs the background is represented by the 
percentage of local charge conservation (LCC) within an event. This 
number can be considered as the amount of positive and negative charged 
partners emitted from the same fluid element relative to the total 
multiplicity of the event.

The first step in the whole procedure was to calibrate the model 
without the inclusion of any CME or LCC effects, in what will be 
referred to in the rest of the text as ``baseline''. This involved 
tuning the input parameters to describe the centrality dependence 
of bulk measurements, such as the charged particle multiplicity density 
$dN/d\eta$~\cite{Abbas:2013bpa,Adam:2015ptt,Acharya:2018hhy} and 
$v_2$~\cite{Aamodt:2010pa,Adam:2016izf,Acharya:2018ihu} in Pb--Pb 
and Xe--Xe collisions at various LHC energies. Overall the model 
was able to describe the experimental measurements within 15\%. 
Finally, we also checked that the slopes of the transverse momentum 
($p_T$) spectra of pions, kaons and protons, in the baseline sample 
of AVFD have a similar centrality dependence as the one reported by 
ALICE in Refs.~\cite{Abelev:2013vea,Acharya:2019yoi,Acharya:2021ljw}. 

\begin{figure}[!h]
    \begin{center}
    \includegraphics[width = 0.5\textwidth]{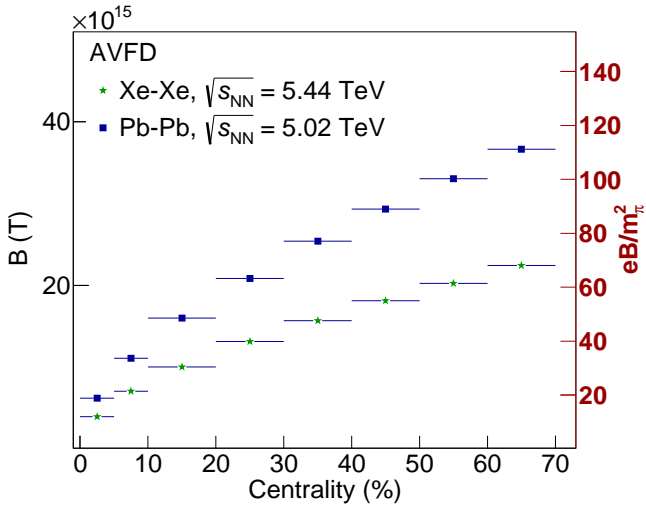}
    \end{center}
    \caption{The dependence of the average value of the magnetic 
    field perpendicular to the reaction plane ($\textbf{B}_y$) on centrality for Pb--Pb and Xe--Xe collisions at 
    $\sqrt{s_{\mathrm{NN}}} = 5.02$~and 
    $\sqrt{s_{\mathrm{NN}}} = 5.44$~TeV, respectively.}
    \label{fig:BFieldVsCentrality}
\end{figure}

One of the key ingredients in the development of the CME in the final 
state, is the early stage electromagnetic field. The 
AVFD model performs an event-by-event simulation of the electromagnetic 
field value projected along the symmetry plane, accounting for the 
decorrelation between the field direction and the true reaction plane 
due to fluctuations~\cite{Shi:2017cpu}. The initial strength of this 
field mainly depends on the atomic number of the nuclei that collide 
and the center-of-mass energy of the collision. Figure~\ref{fig:BFieldVsCentrality} 
presents the centrality dependence of the magnitude of $\textbf{B}$, 
as simulated by AVFD at $t_0 = 0$, for Pb--Pb and Xe--Xe collisions at 
$\sqrt{s_{\mathrm{NN}}} = 5.02$~and $\sqrt{s_{\mathrm{NN}}} = 5.44$~TeV, 
respectively. The values for both systems reach and for some centralities 
even exceed $10^{16}$~T. In addition, the magnitude of $\textbf{B}$ for 
a given centrality interval in collisions between Pb-ions is larger than 
the corresponding value in Xe--Xe collisions by a factor which reflects 
the ratio of the atomic numbers of the two nuclei. 

\begin{figure}[!h]
    \begin{center}
    \includegraphics[width = 0.5\textwidth]{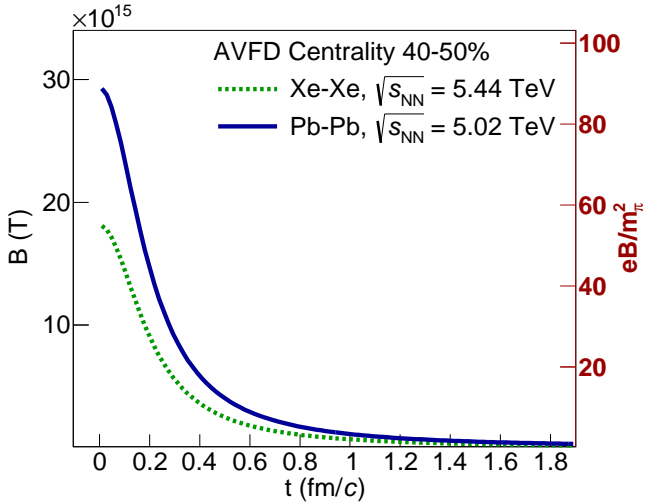}
    \end{center}
    \caption{The time evolution of the average value of the magnetic field 
    perpendicular to the reaction plane ($\textbf{B}$) for the 40\%-50\% 
    centrality interval in Pb--Pb and Xe--Xe collisions at 
    $\sqrt{s_{\mathrm{NN}}} = 5.02$~and $\sqrt{s_{\mathrm{NN}}} = 5.44$~TeV, 
    respectively.}
    \label{fig:BFieldVsTime}
\end{figure}

The magnitude of the field evolves as a function of time in the model 
according to

\begin{equation}
    B(\tau,x)=\frac{B_0}{1+\tau^2/\tau_B^2},
\end{equation}
where $\tau_B$ is the magnetic field lifetime which is set, in this work, 
conservatively to 0.2~fm/$c$, for both collision systems. Figure~\ref{fig:BFieldVsTime} 
presents the time evolution of the magnitude of $\textbf{B}$ for an 
indicative centrality interval i.e. 40-50\% for both Pb--Pb (solid 
line) and Xe--Xe collisions (dashed line).

The next step in the calibration of the model required extracting 
the dependence of the correlators $\Delta \gamma_{1,1}$ based on 
Eq.~\ref{Eq:3ParticleCorrelator} and $\Delta \delta_1$ (see 
Eq.~\ref{Eq:2ParticleCorrelator}) on both the axial current density 
$n_5/\mathrm{s}$ and the percentage of LCC. For this, new AVFD 
samples were produced for all centralities of both systems and 
energies, for which the amount of CME induced signal was 
incremented i.e., using $n_5/\mathrm{s}$ = 0.05, 0.07 and 0.1, 
while at the same time keeping the percentage of LCC fixed at zero. 
In addition, to gauge the dependence of both correlators on the 
background, similar number of events as before were produced where, 
this time, $n_5/\mathrm{s}$ was fixed at zero but the percentage 
of LCC was incremented every time. In particular, the values selected for the Pb-system were 33 and 50\%\footnote{Other values of LCC percentage were also checked, but due to technical reasons related to computing resources, not for all centrality intervals}.

%In particular, the values selected for the Pb-system were 33 and 50\%, while for Xe--Xe collisions events with LCC percentage of 15 and 30\% were studied. The reason for the different values in the two system is technical and is related to computing resources.

\begin{figure}[!h]
    \begin{center}
    \includegraphics[width = 0.5\textwidth]{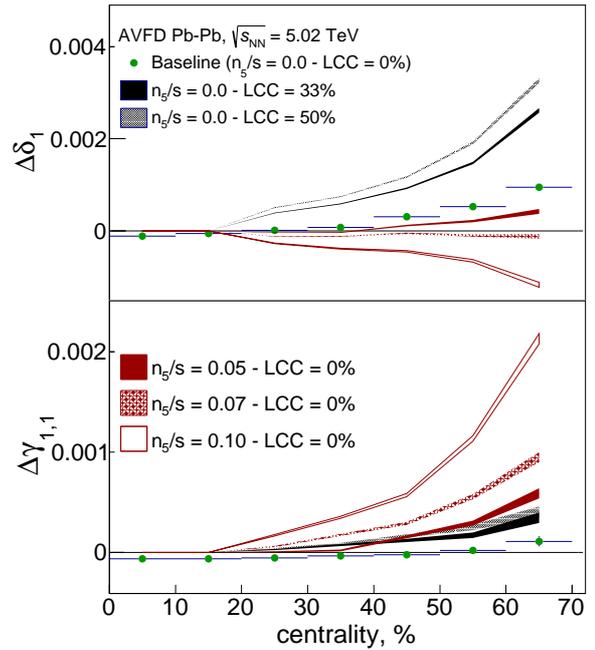}
    \end{center}
    \caption{The centrality dependence of $\Delta \delta_1$ (upper 
    panel) and $\Delta \gamma_{1,1}$ (lower panel), the charge dependent 
    difference of $\delta_1$ and $\gamma_{1,1}$ between opposite- 
    and same-sign pairs, in Pb--Pb collisions at $\sqrt{s_{\mathrm{NN}}} = 5.02$~TeV. 
    The results from the analysis of the baseline sample are 
    represented by the green markers. The various bands show the 
    AVFD expectations for $\Delta \delta_1$ and $\Delta \gamma_{1,1}$ 
    for various values of $n_5/\mathrm{s}$ (red bands) and percentage 
    of LCC (blue bands).}
    \label{fig:ModelCalibPbPb}
\end{figure}

\begin{figure}[!h]
    \begin{center}
    \includegraphics[width = 0.5\textwidth]{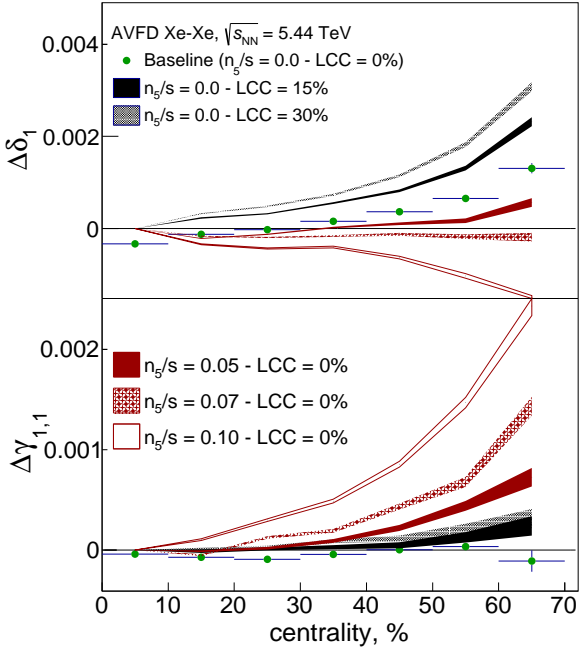}
    \end{center}
    \caption{The centrality dependence of $\Delta \delta_1$ (upper 
    panel) and $\Delta \gamma_{1,1}$ (lower panel), the charge dependent 
    difference of $\delta_1$ and $\gamma_{1,1}$ between opposite- 
    and same-sign pairs, in Xe--Xe collisions at $\sqrt{s_{\mathrm{NN}}} = 5.44$~TeV. 
    The results from the analysis of the baseline sample 
    are represented by the green markers. The various bands show the 
    AVFD expectations for $\Delta \delta_1$ and $\Delta \gamma_{1,1}$ 
    for various values of $n_5/\mathrm{s}$ (red bands) and percentage 
    of LCC (blue bands).}
    \label{fig:ModelCalibXeXe}
\end{figure}

Figure~\ref{fig:ModelCalibPbPb} presents the centrality dependence 
of $\Delta \delta_1$ and $\Delta \gamma_{1,1}$, 
in the upper and lower panels, respectively. The plots show results 
obtained from the analysis of events of Pb--Pb collisions at 
$\sqrt{s_{\mathrm{NN}}} = 5.02$~TeV. 
%Similar plot was extracted for Pb--Pb collisions at $\sqrt{s_{\mathrm{NN}}} = 2.76$~TeV. 
The green markers are extracted from the analysis of the baseline 
sample and, in both cases, exhibit non-zero values for the majority 
of the centrality intervals. These non-zero values are due to the 
existence of hadronic resonances in the model whose decay products 
are affected by both radial and elliptic flow. In addition, the same 
plots present how the magnitude of these correlators develop for 
various values of the axial current density $n_5/\mathrm{s}$ which 
are represented by the red bands. It can be seen that with increasing 
values of $n_5/\mathrm{s}$ the two correlators exhibit opposite 
trends: while $\Delta \delta_1$ decreases, the values of 
$\Delta \gamma_{1,1}$ increase. This opposite behaviour originates 
from the different sign the CME contributes to $\delta_1$ (Eq.~\ref{Eq:2ParticleCorrelator}) and 
$\gamma_{1,1}$ (Eq.~\ref{Eq:3ParticleCorrelator}) and, consequently, to $\Delta \delta_1$ and 
$\Delta \gamma_{1,1}$. Finally, when fixing 
the value of $n_5/\mathrm{s}$ to zero and progressively increasing 
the percentage 
of LCC in the sample (black curves in fig.~\ref{fig:ModelCalibPbPb}), 
the values of both $\Delta \delta_1$ and $\Delta \gamma_{1,1}$ 
increase. However, the latter correlator exhibits a smaller sensitivity 
than $\Delta \delta_1$ to the background owning to the fact that 
it is constructed as the difference in the magnitude of background 
effects in- and out-of-plane (see Eq.~\ref{Eq:3ParticleCorrelator}).

Similarly, fig.~\ref{fig:ModelCalibXeXe} presents the centrality 
dependence of $\Delta \delta_1$ and $\Delta \gamma_{1,1}$, this 
time in Xe--Xe collisions at $\sqrt{s_{\mathrm{NN}}} = 5.44$~TeV. 
Also here the results for the baseline AVFD sample are represented 
with the green markers, while the red and black bands correspond 
to samples with progressively increasing values of $n_5/\mathrm{s}$ 
and percentage of LCC, respectively. The same qualitative observations 
are also found in this system: the baseline sample exhibits non-zero 
values for both $\Delta \delta_1$ and $\Delta \gamma_{1,1}$, these 
two correlators have opposite trends with increasing $n_5/\mathrm{s}$ 
and $\Delta \delta_1$ exhibits bigger sensitivity on the LCC percentage 
than $\Delta \gamma_{1,1}$.

\begin{figure}[!h]
    \begin{center}
    \includegraphics[width = 0.5\textwidth]{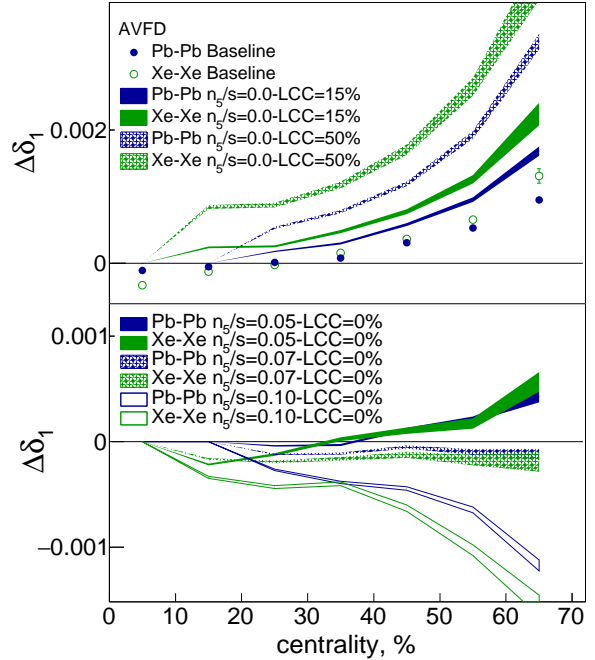}
    \end{center}
    \caption{The centrality dependence of $\Delta \delta_1$  grouped in two scenarios: zero $n_5/\mathrm{s}$ but various choices of LCC (upper panel) and non-zero $n_5/\mathrm{s}$ but LCC fixed to zero (lower panel). The various bands show the AVFD expectations for $\Delta \delta_1$ 
    for Pb--Pb collisions and Xe--Xe collisions, with blue and green bands, respectively. The results of the baseline sample are represented by the filled and open markers.}
    \label{fig:CompareDeltaD11Pb5andXe}
\end{figure}

\begin{figure}[!h]
    \begin{center}
    \includegraphics[width = 0.5\textwidth]{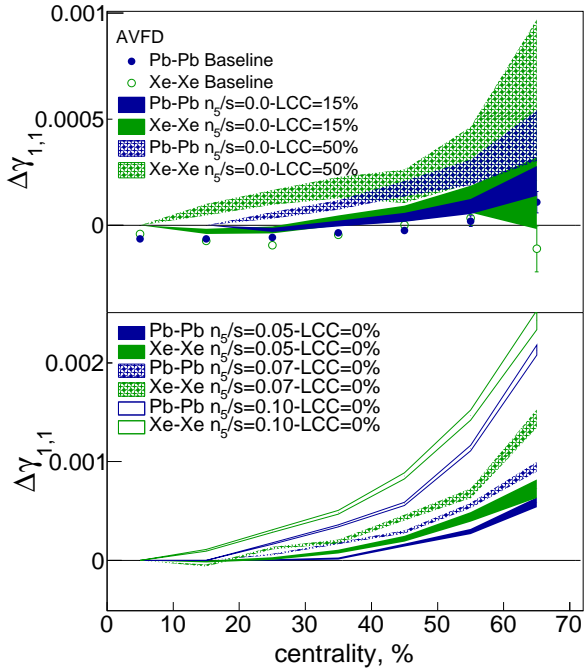}
    \end{center}
    \caption{The centrality dependence of $\Delta \gamma_{1,1}$  grouped in two scenarios: zero $n_5/\mathrm{s}$ but various choices of LCC (upper panel) and non-zero $n_5/\mathrm{s}$ but LCC fixed to zero (lower panel). The various bands show the AVFD expectations for $\Delta \delta_1$ 
    for Pb--Pb collisions and Xe--Xe collisions, with blue and green bands, respectively. The results of the baseline sample are represented by the filled and open markers.}
    \label{fig:CompareDeltaG112Pb5andXe}
\end{figure}

To directly compare the values of $\Delta\delta_1$ and $\Delta\gamma_{1,1}$ between these two collision systems, the results are organised based on the input parameters used: zero $n_5/\mathrm{s}$ but various choices of LCC and non-zero $n_5/\mathrm{s}$ but LCC fixed to zero. 
Figures~\ref{fig:CompareDeltaD11Pb5andXe} and~\ref{fig:CompareDeltaG112Pb5andXe}, summarize the centrality dependence of the results for $\Delta\delta_1$ and $\Delta\gamma_{1,1}$. 

In the first case, the baseline and LCC being 15\% and 50\% for Pb--Pb collisions at $\sqrt{s_{NN}}=5.02$ TeV and for Xe--Xe collisions at $\sqrt{s_{NN}}=5.44$ TeV are chosen. The upper panel of fig.~\ref{fig:CompareDeltaD11Pb5andXe} illustrates that for a fixed LCC percentage, the values of $\Delta\delta_1$ are higher for the Xe--Xe than for the Pb--Pb samples. For a fixed centrality, while the effect of radial flow between these two systems is similar~\cite{Acharya:2021ljw}, the charged particle multiplicity in Pb--Pb is 60--70\% higher than the corresponding value in Xe--Xe collisions~\cite{Adam:2015ptt,Acharya:2018hhy}. This could lead to a faster dilution of the correlations induced by the LCC mechanism in the larger system, reflected in this difference of $\Delta\delta_1$. At the same time, the upper panel of fig.~\ref{fig:CompareDeltaG112Pb5andXe} shows that the values of $\Delta\gamma_{1,1}$ for the two systems do not exhibit any significant difference. This is in line with the expectation that the sensitivity of $\Delta\gamma_{1,1}$ to the background is significantly reduced with respect to $\Delta \delta_1$.

In the second case of non-zero axial current density, the samples containing $n_5/\mathrm{s}=0.05, 0.07$ and $0.10$ are chosen. The lower panel of fig.~\ref{fig:CompareDeltaD11Pb5andXe} shows that $\Delta\delta_1$ is similar between the two systems since it is primarily affected by background contributions. This correlator needs higher values of $n_5/\mathrm{s}$ (e.g. $n_5/\mathrm{s} = 0.1$ in the plot) to start observing some differences. Finally, the lower panel of fig.~\ref{fig:CompareDeltaG112Pb5andXe} illustrates that the magnitude of $\Delta\gamma_{1,1}$ is higher in the Xe--Xe than in the Pb--Pb samples. Although the value of the magnetic field is higher for the larger Pb-system, as shown in fig.~\ref{fig:BFieldVsCentrality}, the significantly larger multiplicity that this system has, leads to a larger dilution effect reflected in the ordering of the corresponding curves in the plot.

%Since the magnetic field is higher for Pb-Pb collisions mainly due to higher atomic number, as shown in fig.~\ref{fig:BFieldVsCentrality}, the corresponding magnitude of the CME component, denoted by $\langle a_{1,\alpha}a_{1,\beta}\rangle$, is higher as a consequence. According to Eq.~\ref{Eq:3ParticleCorrelator} and Eq.~\ref{Eq:2ParticleCorrelator}, the CME component has a positive contribution to $\Delta\delta_1$ and a negative contribution to $\Delta\gamma_{1,1}$, leading to the observed behaviour in the second case.

\begin{figure}[!h]
    \begin{center}
    \includegraphics[width = 0.5\textwidth]{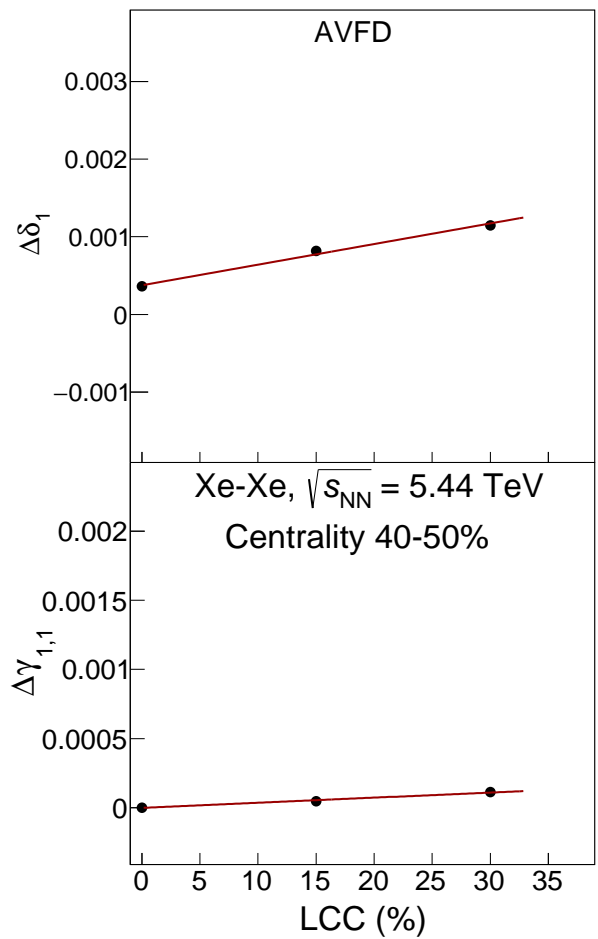}
    \end{center}
    \caption{The dependence of $\Delta \delta_1$ and $\Delta \gamma_{1,1}$ 
    in the upper and lower panels, respectively, on the percentage of local 
    charge conservation in the analysed samples of Pb--Pb collisions at 
    $\sqrt{s_{\mathrm{NN}}} = 5.02$~TeV for the 40\%-50\% centrality interval.}
    \label{fig:ResultsVsLCC}
\end{figure}

\begin{figure}[!h]
    \begin{center}
    \includegraphics[width = 0.5\textwidth]{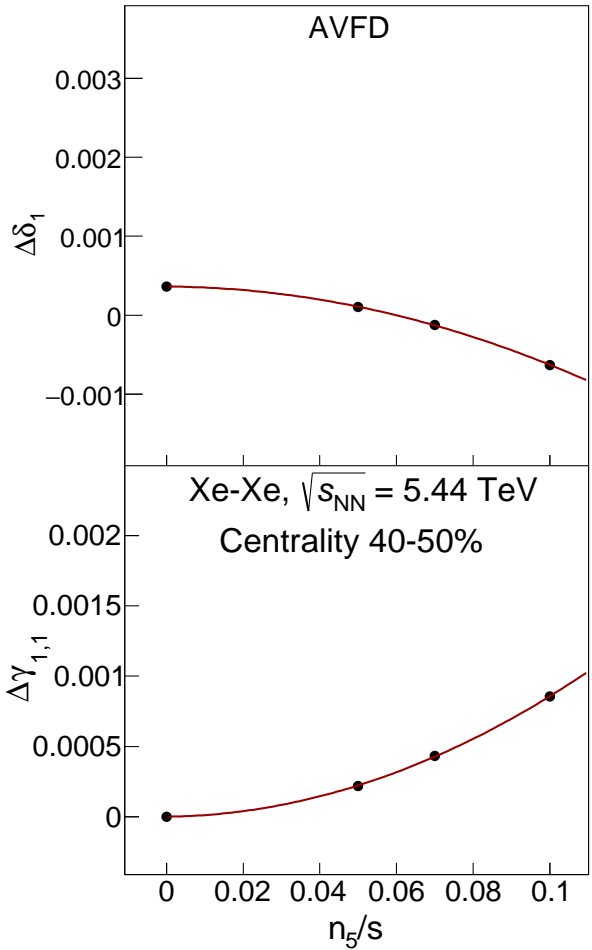}
    \end{center}
    \caption{The dependence of $\Delta \delta_1$ and $\Delta \gamma_{1,1}$ 
    in the upper and lower panels, respectively, on the axial current density 
    $n_5/\mathrm{s}$ in the analysed samples of Pb--Pb collisions at 
    $\sqrt{s_{\mathrm{NN}}} = 5.02$~TeV for the 40\%-50\% centrality interval.}
    \label{fig:ResultsVsCME}
\end{figure}

The previous results for each colliding system and energy can 
be grouped in a different way that allows to parametrise the 
dependence of each of the correlators on the LCC percentage 
and on $n_5/\mathrm{s}$. Figures~\ref{fig:ResultsVsLCC} 
and~\ref{fig:ResultsVsCME} present how $\Delta \delta_1$ and 
$\Delta \gamma_{1,1}$ develop as a function of the LCC percentage 
and $n_5/\mathrm{s}$, respectively. Results for the 40--50\% 
centrality interval of Pb--Pb collisions at 
$\sqrt{s_{\mathrm{NN}}} = 5.02$~TeV are indicatively chosen to 
illustrate the procedure. An identical protocol was used for all 
centrality intervals of both colliding systems. One can see that 
both $\Delta \delta_1$ and $\Delta \gamma_{1,1}$ exhibit a linear 
dependence on the percentage of LCC, with the latter being less 
sensitive and thus having a smaller slope. Finally, these two 
correlators exhibit a quadratic dependence on $n_5/\mathrm{s}$ 
with opposite trend, originating from the dependence of $\delta_1$ 
and $\gamma_{1,1}$ on $\langle a_{1,\alpha} a_{1,\beta}\rangle$ and 
$-\langle a_{1,\alpha} a_{1,\beta}\rangle$ in Eq.~\ref{Eq:2ParticleCorrelator} 
and Eq.~\ref{Eq:3ParticleCorrelator}, respectively. This $a_1$ 
coefficient, in turns, has been shown in Ref.~\cite{Shi:2017cpu,Jiang:2016wve} 
to be proportional to the value of $n_5/\mathrm{s}$.

Following this procedure for all centrality intervals of Pb--Pb and 
Xe--Xe collisions, one is able to parametrise the dependence of 
$\Delta \delta_1$ and $\Delta \gamma_{1,1}$ according to:

\begin{equation}
\Delta \delta_1 = c_2 \cdot (n_5/\mathrm{s})^2 + c_1 \cdot (n_5/\mathrm{s}) + b_1 \cdot (\mathrm{LCC}) + b_0,
\label{Eq:DeltaParametrisation}
\end{equation}

\begin{equation}
\Delta \gamma_{1,1} = e_2 \cdot (n_5/\mathrm{s})^2 + e_1 \cdot (n_5/\mathrm{s}) + d_1 \cdot (\mathrm{LCC}) + d_0,
\label{Eq:GammaParametrisation}
\end{equation}

\noindent where $e_2$, $e_1$, $d_1$, $d_0$, $c_2$, $c_1$, $b_1$ and $b_0$ 
are real numbers constrained from the simultaneous fit of the corresponding 
dependencies of $\Delta \delta_1$ and $\Delta \gamma_{1,1}$ for each centrality 
interval of every collision system and energy. The parametrisation of 
Eq.~\ref{Eq:DeltaParametrisation} and Eq.~\ref{Eq:GammaParametrisation} 
assumes that the two components that control the CME signal and the background 
are not correlated. This is a reasonable assumption considering that the 
two underlying physical mechanism are independent and take place at different 
times in the evolution of a heavy ion collision.

\section{Results}
\label{Sec:Results}

Having the dependence of both $\Delta \delta_1$ and 
$\Delta \gamma_{1,1}$ on $n_5/\mathrm{s}$ and LCC parametrised 
from Eq.~\ref{Eq:DeltaParametrisation} and 
Eq.~\ref{Eq:GammaParametrisation}, one can deduce the values of 
these two parameters that govern the CME signal and the background 
for each centrality, colliding system and energy that allows, at 
the same time, for a quantitative description of the measured 
centrality dependence of $\Delta \delta_1$ and $\Delta \gamma_{1,1}$ 
at LHC energies.

\begin{figure}[!h]
    \begin{center}
    \includegraphics[width = 0.5\textwidth]{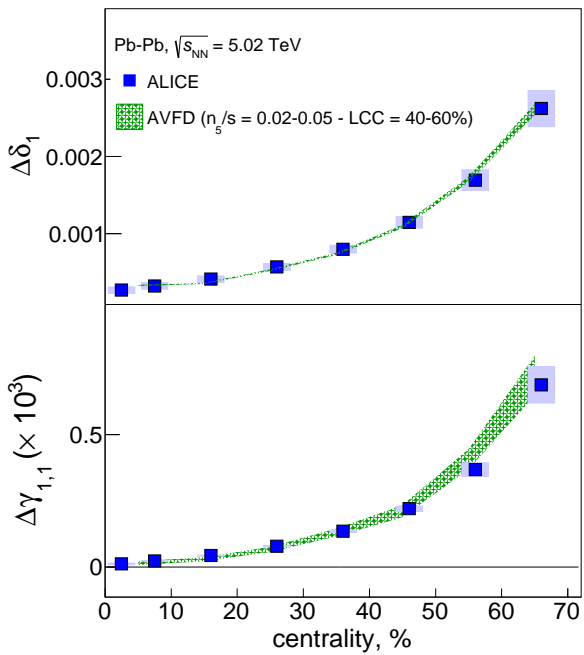}
    \end{center}
    \caption{The centrality dependence of $\Delta \delta_1$ and 
    $\Delta \gamma_{1,1}$ in the upper and lower panels, respectively. 
    The data points represent the experimental measurements in Pb--Pb collisions at $\sqrt{s_{\mathrm{NN}}} = 5.02$~TeV~\cite{Acharya:2020rlz}. The green band shows the results 
    obtained from the tuned AVFD sample (see text for details).}
    \label{fig:DataVsModel}
\end{figure}

Figure~\ref{fig:DataVsModel} presents the results of such procedure 
for Pb--Pb collisions at $\sqrt{s_{\mathrm{NN}}} = 5.02$~TeV. The 
data points, extracted from Ref.~\cite{Acharya:2020rlz} for both 
correlators are described fairly well by the tuned model. A similarly 
satisfactory description is also achieved for the results 
%of Pb--Pb collisions at $\sqrt{s_{\mathrm{NN}}} = 2.76$~TeV~\cite{Abelev:2012pa} and for the relevant measurements in 
of Xe--Xe collisions at 
$\sqrt{s_{\mathrm{NN}}} = 5.44$~TeV~\cite{Aziz:2020nia}.

\begin{figure}[!h]
    \begin{center}
    \includegraphics[width = 0.5\textwidth]{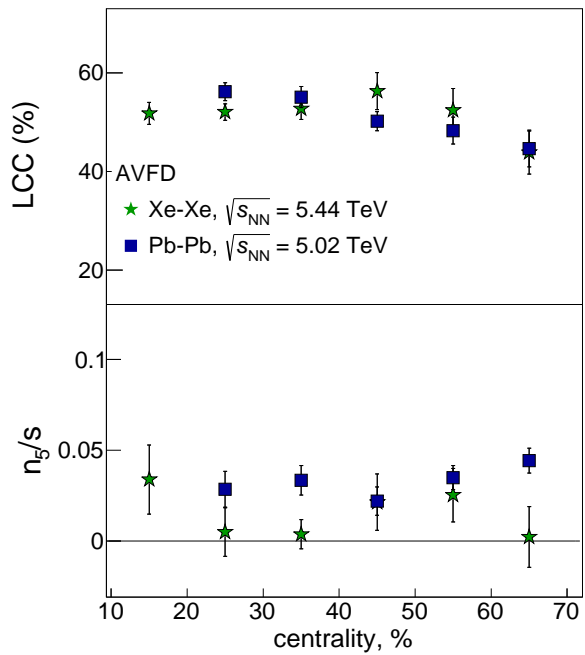}
    \end{center}
    \caption{The centrality dependence of the LCC percentage (upper 
    panel) and the axial current density $n_5/\mathrm{s}$ that allows to describe simultaneously the experimental measurements of $\Delta \delta_1$ 
    and $\Delta \gamma_{1,1}$~\cite{Abelev:2012pa,Acharya:2020rlz,Aziz:2020nia} 
    in all collision systems and energy studied in this article.}
    \label{fig:ModelResults}
\end{figure}

Figure~\ref{fig:ModelResults} presents the final result of the whole 
procedure. The plots show the centrality dependence of the pairs of 
LCC percentage (upper panel) and $n_5/\mathrm{s}$ (lower panel) that 
are needed to describe with AVFD the experimental measurements of 
$\Delta \delta_1$ and $\Delta \gamma_{1,1}$. The different markers 
represent results for different collision systems and energies. It 
can be seen that all systems can be described by large values of LCC 
that range from 40$\%$ for peripheral up to around 60$\%$ for more 
central Pb--Pb collisions. There is no significant difference observed 
in these values among the two sets of results. 

Furthermore, the lower panel of fig.~\ref{fig:ModelResults} 
illustrates that there is no significant centrality dependence 
of $n_5/\mathrm{s}$. However, there is a dependence on the colliding 
system. More particularly, the experimental results from the 
analysis of Xe--Xe collisions lead to values of $n_5/\mathrm{s}$ 
which are compatible with zero within the uncertainties for all 
centrality intervals. A fit with a constant function results 
into values of $0.011 \pm 0.005$. At the same time, the results 
for Pb--Pb collisions can be described by non-zero values of axial 
current densities, again for the entire centrality region studied. 
The corresponding fit leads to a value of $0.034 \pm 0.003$ i.e., 
significantly above the background-only scenario.

\section{Summary}
\label{Sec:Summary}
In this article we presented a systematic study of charge 
dependent azimuthal correlations which are commonly used 
experimentally to probe the Chiral Magnetic Effect using 
the Anomalous-Viscous Fluid Dynamics 
framework~\cite{Shi:2017cpu,Jiang:2016wve}. After tuning 
the model to reproduce, within 15\%, basic experimental 
measurements such 
as the centrality dependence of the charged particle multiplicity 
density and the elliptic flow we were able to parametrise 
the dependence of both $\Delta \delta_1$ and $\Delta \gamma_{1,1}$ 
on the LCC percentage, the main contribution to the background, 
and the axial current density $n_5/\mathrm{s}$ which dictates 
the amount of CME signal. This procedure was followed for Pb--Pb 
collisions at $\sqrt{s_{\mathrm{NN}}} = 5.02$~TeV, as well as 
for Xe--Xe collisions at $\sqrt{s_{\mathrm{NN}}} = 5.44$~TeV. 
This parametrisation allowed for the estimation of the values 
of both the LCC percentage and $n_5/\mathrm{s}$ needed to describe 
quantitatively and at the same time the centrality dependence 
of $\Delta \delta_1$ and $\Delta \gamma_{1,1}$ measured 
experimentally~\cite{Abelev:2012pa,Acharya:2020rlz,Aziz:2020nia}. 
The measurements in Xe--Xe are consistent with a background only 
scenario, with values of $n_5/\mathrm{s}$ compatible with zero. 
On the other hand, the results of Pb--Pb collisions require 
$n_5/\mathrm{s}$ with significantly non-zero values.

%------------------------------------------------%

%------------------------------------------------%
% For one-column wide figures use
%\begin{figure}
%    \includegraphics{example.eps}
%    \caption{Please write your figure caption here}
%    \label{fig:1}       % Give a unique label
%\end{figure}
%
% For two-column wide figures use
%\begin{figure*}
%    \includegraphics[width=0.75\textwidth]{example.eps}
%    \caption{Please write your figure caption here}
%    \label{fig:2}       % Give a unique label
%\end{figure*}
%------------------------------------------------%

\begin{acknowledgements}
We are grateful to Prof.~Jinfeng Liao and Dr.~Shuzhe Shi for 
providing the source code of the model, for their guidance 
and their feedback during this study. We would like to thank 
Prof. Sergei Voloshin for his suggestions.

\end{acknowledgements}

% BibTeX users please use one of
%\bibliographystyle{spbasic}      % basic style, author-year citations
%\bibliographystyle{spmpsci}      % mathematics and physical sciences
%\bibliographystyle{spphys}       % APS-like style for physics
%\bibliography{}   % name your BibTeX data base
\bibliographystyle{utphys}
\bibliography{cmeModelPaperEPJC.bib}{}

\end{document}